\documentclass[a4paper, 12pt]{article} 
\usepackage[right=1.1in,left=1.1in,top=1.1in,bottom=1.1in]{geometry}
\usepackage{graphicx} 
\usepackage{braket}
\usepackage{mathrsfs}
\usepackage[T1]{fontenc} 
\usepackage{amsmath}
\usepackage{xypic}

\usepackage{pxfonts}
\usepackage[T1]{fontenc}
\usepackage{amssymb}
\usepackage{natbib}
\usepackage{epstopdf}
\usepackage{setspace}
\usepackage{enumitem}
\usepackage{sectsty}
\usepackage{wrapfig} 
\sectionfont{\large}
\bibpunct{(}{)}{,}{a}{}{,}
\usepackage{tikz}   
\usepackage{tikz-3dplot} 

\newtheorem{theorem}{Theorem}[section]

\newtheorem{corollary}[theorem]{Corollary}

\newcommand{\qed}{\nobreak \ifvmode \relax \else
      \ifdim\lastskip<1.5em \hskip-\lastskip
      \hskip1.5em plus0em minus0.5em \fi \nobreak
      \vrule height0.75em width0.5em depth0.25em\fi}

\usepackage{bbm}
\usepackage{MnSymbol}
\usepackage[all]{xy}

\usepackage{xcolor,colortbl,array,amssymb}

   \usepackage{braket}
   \usepackage{amssymb}
\usepackage{multicol}

\makeatletter

\title{ \large{ Quantum States of a Time-Asymmetric Universe:  }
\\ \large{Wave Function, Density Matrix, and Empirical Equivalence}
} 

\author{Eddy Keming Chen\thanks{Department of Philosophy, 106 Somerset Street, Rutgers University, New Brunswick, NJ 08901, USA. Website: www.eddykemingchen.net. Email: eddy.chen@rutgers.edu  }}  

\date{\today} 

\begin{document}
\bibliographystyle{apalike}

\maketitle 



\begin{abstract}

What is the quantum state of the universe? Although there have been several interesting suggestions, the question remains open. In this paper, I consider a natural choice for the universal quantum state  arising from the Past Hypothesis, a boundary condition that accounts for the time-asymmetry of the universe. The natural choice is  given not by a wave function (representing a pure state) but by a density matrix (representing a mixed state).  

 I  begin  by classifying quantum theories into two types: theories with a fundamental wave function and theories with a fundamental density matrix. The Past Hypothesis is compatible with infinitely many initial wave functions, none of which seems to be particularly natural. However, once we turn to density matrices, the Past Hypothesis  provides a natural choice---the normalized projection onto the Past Hypothesis subspace in the Hilbert space. Nevertheless, the two types of theories can be empirically equivalent. To provide a concrete understanding of the empirical equivalence, I provide a novel subsystem analysis in the context of Bohmian theories. Given the empirical equivalence, it seems empirically underdetermined whether the universe is in a pure state or a mixed state. Finally, I discuss some theoretical payoffs of the density-matrix theories and  present some open problems for future research.

\end{abstract}

\hspace*{3,6mm}\textit{Keywords: time's arrow, Past Hypothesis, Initial Projection Hypothesis, Statistical Postulate, typicality, unification, foundations of probability, quantum statistical mechanics, wave function realism,  density matrix, quantum state of the universe}   

\newpage

\begingroup
\singlespacing
\tableofcontents
\endgroup

\vspace{20pt} 



\nocite{ AlbertLPT, albert2000time, loewer2004david, lebowitz2008time, goldstein2001boltzmann, durr1992quantum, durr2012quantum, goldstein2013reality, goldstein2010approachB, goldstein2010approach, goldstein2010normal, durr2005role, bell1980broglie, goldstein2012typicality, ney2013wave, ChenOurFund, ChenHSWF, chen2017intrinsic, allori2013primitive, allori2008common, loewer2016mentaculus, LewisPP2, sep-time-thermo, north2011time}

\section{Introduction}

What is the quantum state of the universe? There are probably many universal quantum states compatible with our observations. But is there a particularly natural choice? There have been several interesting suggestions. In the context of quantum gravity, \cite{hartle1983wave} propose that the quantum state of the universe is the natural wave function of a universe with ``no boundaries.'' It is given by a path integral over all compact Euclidean four-geometries which have the relevant three-geometry and its matter configuration at the boundary. This pure state is positive and real; it also satisfies the Wheeler-DeWitt equation. 

In this paper, we turn to quantum statistical mechanics and consider another natural choice for the universal quantum state. It arises from considerations about the Past Hypothesis, which is an attempt to account for the thermodynamic arrow of time in our universe. The natural choice is not a pure state, represented by a wave function, but a mixed state, represented by a density matrix. Moreover, it is complex-valued (in the case of spinless particles). This offers a perspective that may be complementary to the proposal of \cite{hartle1983wave}. 

Since the goal of this paper is largely conceptual, we will focus on the simple context of non-relativistic quantum mechanics and leave to future work how to extend this proposal to more advanced theories such as quantum gravity. In the simple context, we will consider six different theories that fall into two types: 

\begin{enumerate}
\item Theories with a fundamental wave function.
\item Theories with a fundamental density matrix. 
\end{enumerate}
The qualification ``fundamental'' is necessary. First, in a universe with a pure state, impure density matrices can nonetheless emerge both as  descriptions of our ignorance (statistical density matrices) and as subsystem descriptions (reduced density matrices).  Second, in a universe with a mixed state, wave functions can emerge at the subsystem level. 

In \S2, we  begin by formulating three theories in each type and consider how to combine them with the Past Hypothesis. We notice that there is an interesting difference. In theories with a fundamental wave function, there is a canonical measure of probability (or typicality) but no natural choice of the initial quantum state. In theories with a fundamental density matrix, there is a natural choice of the initial quantum state but there does not seem to be a canonical measure of probability (or typicality). Hence, given the Past Hypothesis, only in the second type of theories do we obtain a natural choice for the universal quantum state.  For each type, we also discuss how to combine it with solutions to the quantum measurement problem: Bohmian mechanics, Everettian  mechanics, and GRW spontaneous collapse theories. 

In \S3, we suggest that the two types of theories can nonetheless be made empirically equivalent. In particular, if the probability measure over wave functions gives rise to a statistical density matrix that equals to the fundamental density matrix in the other theory, the two theories will have the same  probability distributions over measurement outcomes. In  Bohmian theories, this fact lets us prove a general theorem about empirical equivalence, from which we obtain an important corollary. The general argument from equivalent probability distributions will be supplemented by a novel analysis about the Bohmian subsystems. Empirical equivalence also apply to the two types of Everettian theories and the two types of GRW theories. 

In \S4, we discuss some theoretical payoffs of the density-matrix theories in comparisons with the wave-function theories. In particular, the density-matrix theories  lead to a simple and (more or less) unique initial quantum state. It makes the nomological interpretation of the quantum state much more compelling, reduces statistical mechanical probabilities to quantum mechanical ones, and provides significant gain in theoretical unification. There are also interesting open problems  in the density-matrix approach. A noteworthy consequence is that our particular choice of the initial quantum state will  be non-normalizable if the Past Hypothesis subspace of the Hilbert space turns out to be infinite-dimensional. This requires us to face head-on the problem of non-normalizable quantum states, a possibility that has so far been discussed mainly in the context of quantum gravity and quantum cosmology.

\section{Wave Function, Density Matrix, and the Past Hypothesis}

In this section, we first discuss quantum mechanics with a fundamental wave function. Our discussion will be informed by  Boltzmannian quantum statistical mechanics and solutions to the quantum measurement problem. We will then turn to quantum theories with a fundamental density matrix and explain why, given the Past Hypothesis, there is a natural density matrix.

\subsection{Quantum Mechanics with a Fundamental Wave Function}

\subsubsection{$\Psi$-QM}

Standard quantum mechanics is often presented with a set of axioms and rules about measurement. Firstly, there is a quantum state of the system, represented by a wave function $\psi$. For a spin-less $N$-particle quantum system in $\mathbb{R}^3$, the wave function is a (square-integrable) function from the configuration space $\mathbb{R}^{3N}$ to the complex numbers $\mathbb{C}$. Secondly, the wave function evolves deterministically according to the the Schr\"odinger equation:

\begin{equation}\label{SE}
 i\hbar \frac{\partial \psi}{\partial t} = H \psi
\end{equation}
Thirdly, the Schr\"odinger evolution of the wave function is supplemented with collapse rules. The wave function typically evolves into superpositions of macrostates, such as the cat being alive and the cat being dead. This can be represented by wave functions on the configuration space with disjoint macroscopic supports $X$ and $Y$. During measurements, which are not precisely defined processes in the standard formalism, the wave function  undergoes collapses. Moreover,  the probability that it collapses into any particular macrostate $X$ is given by the Born rule: 

\begin{equation}\label{Born}
P( X) = \int_{X} |\psi(x)|^2 dx
\end{equation}

As such, quantum mechanics is not a candidate for a fundamental physical theory. It has two dynamical laws: the deterministic Schr\"odinger equation and the stochastic collapse rule. What are the conditions for applying the former, and what are the conditions for applying the latter? Measurements and observations are extremely vague concepts. Take a concrete experimental apparatus for example. When should we treat it as part of the quantum system that evolves linearly and when should we treat it as an ``observer,'' i.e. something that stands outside the quantum system and collapses the wave function? That is, in short, the quantum measurement problem.\footnote{See \cite{bell1990against} and \cite{sep-qt-issues} for  introductions to the quantum measurement problem.} 

Various solutions have been proposed regarding the measurement problem. Bohmian mechanics (BM) solves it by adding particles to the ontology and an additional guidance equation for the particles' motion. Ghirardi-Rimini-Weber (GRW)  theories postulate a spontaneous collapse mechanism that disrupts the linear Schr\"odinger evolution of the wave function. Everettian quantum mechanics (EQM), according to the ``Oxford interpretation,''\footnote{See \cite{wallace2012emergent} for an up to date development and defense.} simply removes the collapse rules from standard quantum mechanics and suggest that there are many (emergent) worlds, corresponding to the branches of the wave function, which are all real. My aim here is not to adjudicate among these theories. Suffice it to say that they are all quantum theories that remove the centrality of observations and observers. 

In the following, we will write down the standard formalism of BM, GRW, and EQM with a fundamental wave function. 

(1) \textbf{$\Psi$-BM.} In addition to the wave function $\Psi$ that evolves unitarily according to the Schr\"odinger equation,  there are actual particles that have precise locations in physical space, represented by $\mathbb{R}^3$. The particle configuration $Q = (Q_1, Q_2, ... , Q_N) \in \mathbb{R}^{3N}$ follows the guidance equation (written for the $i$-th particle): 
\begin{equation}\label{GE}
 \frac{dQ_i}{dt} = \frac{\hbar}{m_i} \text{Im} \frac{ \nabla_i \psi (q) }{  \psi (q)} (q=Q)
\end{equation}
Moreover, the initial particle distribution is given by the quantum equilibrium distribution: 
\begin{equation}\label{QEH}
\rho_{t_0} (q) = |\psi(q, t_0)|^2
\end{equation}
By equivariance, if this condition holds at the initial time, then it holds at all times.  (See \cite{durr1992quantum}.) Consequently, BM agrees with standard quantum mechanics with respect to the Born rule predictions (which are all there is to the observable predictions of quantum mechanics).  For a universe with $N$ particles, let us call the wave function of the universe the \emph{universal wave function} and denote it by $\Psi(\boldsymbol{q_1, q_2, ... q_N})$. 

(2) \textbf{$\Psi$-GRW.} There are several versions of GRW theories. In the first one, $\Psi$-GRW0, the fundamental ontology consists only in terms of the universal wave function. The wave function typically obeys the Schr\"odinger equation, but the linear evolution is interrupted randomly (with rate $N\lambda$, where $N$ is the number of particles and $\lambda$ is a new constant of nature of order $10^{-15}$ s$^{-1}$) by collapses: 
\begin{equation}\label{WFcollapse}
\Psi_{T^+} = \frac{\Lambda_{I_{k}} (X)^{1/2} \Psi_{T^-} }{||  \Lambda_{I_{k}} (X)^{1/2} \Psi_{T^-}  || },
\end{equation}
where the collapse center $X$ is chosen randomly with probability distribution $\rho(x) = ||  \Lambda_{I_{k}} (x)^{1/2} \Psi_{T^-}  ||^2 dx$,  $I_k \in \{1, 2, ... N\}$ is chosen  randomly with uniform distribution on that set of particle labels, and the collapse rate operator is defined as:
\begin{equation}\label{collapserate}
\Lambda_{I_{k}} (x) = \frac{1}{(2\pi \sigma^2)^{3/2}} e^{-\frac{(Q_k -x)^2}{2\sigma^2}}
\end{equation}
where $Q_k$ is the position operator of ``particle'' $k$, and $\sigma$ is another new constant of nature of order $10^{-7}$ m postulated in current GRW theories. 

It has been argued  that the GRW theory can and should be given a primitive ontology, i.e. fundamental and localized quantities in physical space.\footnote{See, for example, \cite{allori2007fundamental, allori2008common, allori2013primitive} and \cite{bell1995there}.} There are two choices of primitive ontology: mass densities and flashes. In $\Psi$-GRWm, the fundamental ontology includes a mass-density function on physical spacetime:  
\begin{equation}\label{mxt}
m(x,t) = \bra{\Psi(t)} M(x) \ket{\Psi(t)},
\end{equation}
where $x$ is a physical space variable, $M(x) = \sum_i m_i \delta (Q_i - x)$ is the mass-density operator, which is defined via the position operator $Q_i \psi (q_1, q_2, ... q_n)= q_i \psi (q_1, q_2, ... q_n) $. This allows us to determine the mass-density ontology at time $t$  via $\Psi(t)$. In $\Psi$-GRWf, the fundamental ontology includes $F$, the collection of spacetime event, the spatial components of which are the centers of the spontaneous collapses:
\begin{equation}\label{F}
F= \{ (X_1, T_1), (X_2, T_2), ... (X_k, T_k), ...  \}
\end{equation}

(3) \textbf{$\Psi$-EQM.} There are several versions of Everettian theories. In the first one, $\Psi$-EQM0, the fundamental ontology consists only in terms of the universal wave function. The wave function always and exactly obeys the Schr\"odinger equation. 

The Everettian theory can also be given a primitive ontology, such as a mass-density function on physical space in the same way as in GRW theories. By using (\ref{mxt}), we define a mass-density $m(x,t)$ as part of the fundamental ontology and obtain  $\Psi$-EQMm. 

The  Everettian theories faces two challenges: the ontology problem and the probability problem.  First, it is \emph{prima facie} unclear whether the fundamental ontology can adequately describe our world. Since there is no spontaneous collapse, the wave function and the mass-density function will have contributions from every outcome of experiments. It has been argued (\cite{wallace2012emergent}) that decoherence can effectively separate them into branches that do not interfere with each other. Second, since every outcome obtains, it is not clear what the Born rule probability means. It has been suggested that we can use Savage-style decision theory (e.g. \cite{wallace2012emergent}) or self-locating probabilities (e.g. \cite{sebens2016self}) to make sense of the probabilities. Here it is not the place to evaluate these proposals. But it is worth pointing these out because the challenges take on  different forms in the density-matrix versions. 

\subsubsection{$\Psi_{PH}$-QM}

$\Psi$-QM with the solutions of the measurement problem attempt to describe various quantum phenomena in nature and in the laboratories, such as the interference patterns of the double-slit experiment and the Stern-Gerlach experiment. With the exception of $\Psi$-GRW theories, they are all time-reversal invariant. 

However, the behaviors of macroscopic systems with large number of particles are often irreversible; they  display the thermodynamic arrow of time---entropy tends to increase until thermal equilibrium. Not only that, we also find that we are currently not in thermal equilibrium and our observations tell us that our past had even lower entropy. To understand these thermodynamic phenomena, it is helpful to use the tools of statistical mechanics. We will review the standard postulates in the Boltzmannian quantum statistical mechanics, we shall largely follow \cite{goldstein2010approach} and \cite{goldstein2010approachB}. Just as in Boltzmannian classical statistical mechanics, we will invoke notions of microstate, macrostate, energy shell, probability measure, Boltzmann entropy, approach to thermal equilibrium, and a low-entropy boundary condition called the Past Hypothesis. However, in the quantum case, the state space is no longer the classical phase space but the Hilbert space of wave functions. 

In Boltzmannian quantum statistical mechanics, it is standard to take the microstate to be the normalized wave function of the system.\footnote{In Bohmian theories, we have the additional microvariables given by particle configurations. It is not clear whether we should include that as part of the statistical mechanical microstate. However, it might help in the case of superposition of macrostates with distinct entropy.} 
\begin{equation}
\psi(\boldsymbol{q_1}, ..., \boldsymbol{q_N}) \in \mathscr{H}_{total} = L^2 (\mathbb{R}^{3N}, \mathbb{C}^k) \text{ , } \parallel \psi \parallel_{L^2} = 1,
\end{equation}
where $\mathscr{H}_{total} = L^2 (\mathbb{R}^{3N}, \mathbb{C}^k)$ is the total Hilbert space of the system, which is also the state space of the wave functions. A wave function is a (normalized) vector in the Hilbert space. The wave function evolves according to the Schr\"odinger equation (\ref{SE}). We will focus on the physically relevant wave functions that are contained in the energy shell:
\begin{equation}
\mathscr{H} \subseteq \mathscr{H}_{total} \text{ , } \mathscr{H} = \text{span} \{ \phi_\alpha : E_\alpha \in [E, E+\delta E ]  \},
\end{equation}
This is the subspace (of the total Hilbert space) spanned by energy eigenstates $\phi_\alpha$ whose eigenvalues $E_\alpha$ belong to the $[E, E+\delta E]$ range.  Let $D = \text{dim} \mathscr{H}$, the number of energy levels between $E$ and $E+\delta E$.\footnote{If the energy spectrum is discrete, it is necessary to use this fattened interval because otherwise there may not be any energy eigenstates with the exact eigenvalue.}  The measure $\mu_E$ is given by the normalized surface area measure on the unit sphere in the energy subspace $\mathscr{S}(\mathscr{H})$.\footnote{In cases where the energy shell is infinite-dimensional,  we should use a Gaussian measure.} With a choice of macro-variables,\footnote{They need to be suitably ``rounded'' \emph{\`a la} \cite{von1955mathematical})} the energy shell $\mathscr{H}$ can be orthogonally decomposed into macro-spaces:
\begin{equation}
\mathscr{H} = \oplus_\nu \mathscr{H}_\nu \text{ , } \sum_\nu \text{dim}\mathscr{H}_\nu  = D
\end{equation}
Each $\mathscr{H}_\nu$ corresponds to a macrostate (more or less to small ranges of values of macro-variables that we have chosen in advance). Typically, a wave function is in a superposition of macrostates and is not entirely in any one of the macrospaces. However, we can make sense of situations where $\psi$ is (in the Hilbert space norm) very close to a macrostate $\mathscr{H}_\nu$: 
\begin{equation}\label{close}
\bra{\psi} I_{\nu}  \ket{\psi} \approx 1,
\end{equation}
where $I_{\nu}$ is the projection operator onto $\mathscr{H}_{\nu}$. This means that almost all of $\ket{\psi}$ lies in $\mathscr{H}_{\nu}$.\footnote{In the Bohmian theories, the particle configuration may help resolve some of the ambiguities even when the universal wave function is in no particular macrostate. If the universal wave function $\Psi (x,y)$ has Y-supports which are macroscopically distinct and if  $Y$ the actual  configuration of the environment is in one of them, then the effective wave function $\Psi(x, Y)$ can be defined to be in a particular macrostate in the sense of (\ref{close}).  } Typically, there is a dominant macro-space $\mathscr{H}_{eq}$ that has a dimension that  is almost equal to D: 
\begin{equation}
\frac{\text{dim} \mathscr{H}_{eq}}{\text{dim} \mathscr{H}} \approx 1.
\end{equation}
A system with wave function $\psi$ is in equilibrium if the wave function $\psi $ is very close to $\mathscr{H}_{eq}$ in the sense of (\ref{close}):  $\bra{\psi} I_{eq}  \ket{\psi} \approx 1.$ Given the definition of (\ref{close}), it is reasonable to expect that $\mu_E$-most wave functions are in thermal equilibrium.\footnote{Had we required the expectation value to be exactly one, a wave function will have to be entirely contained in $\mathscr{H}_{eq}$ to be in thermal equilibrium. But since that is only a proper subspace of the energy shell, complete containment is extremely atypical.}

The Boltzmann entropy of a quantum-mechanical system with wave function $\psi$ that is very close to a macrostate $\mathscr{H}_\nu$ is given by:
\begin{equation}\label{Boltzmann}
S_B (\psi) = k_B \text{log} (\text{dim} \mathscr{H}_\nu ),
\end{equation}
for which $\psi$ is in the macrostate $\mathscr{H}_\nu$ in the sense of (\ref{close}). Any wave function $\psi_{eq}$ in thermal equilibrium macrostate thus has the maximum entropy: 
\begin{equation}
S_B (\psi_{eq}) = k_B \text{log} (\text{dim} \mathscr{H}_{eq} ) \approx  k_B \text{log} (D),
\end{equation}
where $\mathscr{H}_{eq}$ denotes the equilibrium macrostate. A central task of Boltzmannian quantum statistical mechanics is to establish mathematical results that demonstrate (or suggest) the following conjecture: 
\begin{description}
\item[B-Conjecture:] $\mu_E$-most wave functions in any macrostate  will evolve to states of higher entropy and eventually to thermal equilibrium.
\end{description}
The B-Conjecture is highly plausible because the equilibrium macrostate is almost the entire energy shell, in terms of dimensions. Typically, a random walk in the Hilbert space will evolve a microstate into  subspaces (in the sense of (\ref{close})) of higher dimensions, which correspond to higher entropy.  However, as in the classical case,  the B-Conjecture admits exceptions. Due to time-reversal invariance, we know that there exist infinitely many wave functions that will evolve to lower-entropy states. But we expect them to be atypical with respect to $\mu_E$. Although there are many results that are highly suggestive, no such conjecture has been rigorously proven for realistic physical systems. Nonetheless, it is reasonable to expect that it is true. 

Assuming the B-Conjecture, almost any initial wave functions in non-equilibrium states will be on trajectories towards equilibrium. However, that is only part of the puzzle about time's arrow. Why are we currently out of thermal equilibrium, and why was our past of even lower  entropy? To answer those questions, it is standard to postulate a low-entropy initial condition of the universe, which David Albert calls the \emph{Past Hypothesis}. In the quantum case with a fundamental wave function, we postulate that the Past Hypothesis takes the following form: 
\begin{equation}\label{PH}
\Psi(t_0) \in \mathscr{H}_{PH} \text{ , } \text{dim} \mathscr{H}_{PH} \ll \text{dim}\mathscr{H}_{eq} \approx \text{dim} \mathscr{H}
\end{equation}
where $\mathscr{H}_{PH}$ is the Past Hypothesis macrospace with dimension much smaller than that of the equilibrium macrospace and also much smaller than the current macrospace. Hence, the initial state has very low entropy in the sense of (\ref{Boltzmann}). 

Moreover, we  make the Statistical Postulate that the initial wave functions are distributed randomly according to the (normalized) surface area measure $\mu$ on the unit sphere in $\mathscr{H}_{PH}$:
\begin{equation}\label{SP}
\rho(d \psi)  = \mu(d\psi)
\end{equation}
Given a finite-dimensional subspace, the choice of the surface area measure is natural. It gives rise to a uniform probability distribution on the unit sphere. It is with respect to this measure we can say that $\mu$-most (typical) initial wave functions compatible with the Past Hypothesis will approach thermal equilibrium. This gives rise to a statistical version of the Second Law of Thermodynamics. For technical reasons, we assume that $\mathscr{H}_{PH}$ is finite-dimensional, so that we can use the (normalized) surface area measure on the unit sphere as the typicality measure. It remains an open question in QSM about how to formulate the low-entropy initial condition when the initial macro-space is infinite-dimensional. We will come back to this point in \S4.2. 

The Past Hypothesis and the Statistical Postulate are additional postulates about initial condition and the initial probability distribution, which can be added to the $\Psi$-theories. We will call this new family of theories $\Psi_{PH}$-QM, which consists in the following theories: $\Psi_{PH}$-BM, $\Psi_{PH}$-GRW0, $\Psi_{PH}$-GRWm, $\Psi_{PH}$-GRWf, $\Psi_{PH}$-EQM0, and $\Psi_{PH}$-EQMm. They are solutions of the quantum measurement problem that also have resources to account for the thermodynamic arrow of time. For example, $\Psi_{PH}$-BM is defined as the theory with the state given by $(Q, \Psi)$,  dynamical equations (\ref{SE}) and  (\ref{GE}), boundary condition (\ref{PH}), and initial probability distribution (\ref{QEH}) and  (\ref{SP}).

\subsubsection{$\Psi_{?}$-QM}

Each   $\Psi_{PH}$-theory admits many initial quantum states. Given the Past Hypothesis, the initial wave function is randomly chosen from $\mathscr{H}_{PH}$, the Past Hypothesis subspace. As discussed earlier, there is a natural choice for the probability distribution given by the normalized surface area. Is there also a natural choice for the initial wave function? There does not seem to be one. 

Suppose, in the simplest case, the low-entropy initial condition arises from a compact region in the configuration space such that any wave function in $\mathscr{H}_{PH}$ has to have compact support in that region and zero elsewhere. The most natural function on that compact region will be uniform in that entire region. This function will be discontinuous at the boundary,  which means that it will have non-zero inner product with wave functions of arbitrarily high energy. Such a wave function is unphysical. Suppose we were to make this function continuous and differentiable, then we lose uniqueness, because there are many choices for such a function, none of them is particularly more natural than the others. So there does not appear to be a natural choice for $\Psi_0$ even in the simplest case. It presumably generalizes to more complicated cases. 

Hence, a $\Psi_{PH}$-theory with a natural initial wave function does not seem to exist. As we shall see, the situation is radically transformed in the density-matrix theories. 

\subsection{Quantum Mechanics with a Fundamental Density Matrix}

In this section, we consider quantum mechanics with a fundamental density matrix. The density matrix plays the same dynamical role as the wave function does in the previous theories. 

\subsubsection{$W$-QM}

In $W$-QM, the quantum state of a system is represented by a density matrix $W$. $W$ is the complete characterization of the quantum state; it does not refer to a statistical state representing our ignorance of the underlying wave function.  For a spin-less $N$-particle quantum system,  a density matrix of the system is a positive, bounded, self-adjoint operator $\hat{W} : \mathscr{H} \rightarrow \mathscr{H}$ with $\text{tr} {\hat{W}} = 1$, where $\mathscr{H}$ is the Hilbert space of the system. In terms of the configuration space $\mathbb{R}^{3N}$, the density matrix can be viewed as a function $W: \mathbb{R}^{3N} \times \mathbb{R}^{3N} \rightarrow \mathbb{C}$. 

A density matrix $\hat{W}$ is pure if $\hat{W} = \ket{\psi} \bra{\psi}$ for some $\ket{\psi}$. Otherwise it is mixed. For $W$-QM, the quantum state of a closed system (or that of the universe) can be pure or mixed. In the first instance, $\hat{W}$ always evolves deterministically according to the von Neumann equation: 
\begin{equation}\label{VNM}
i \hbar \frac{d \hat{W}(t)}{d t} = [\hat{H},  \hat{W}].
\end{equation}
Equivalently: 
\begin{equation}\label{VNMQ}
i \hbar \frac{\partial W(q, q', t)}{\partial t} = \hat{H}_q W(q, q', t) -  \hat{H}_{q'} W(q, q', t),
\end{equation}
where  $\hat{H}_{q}$ means that the Hamiltonian  $\hat{H}$ acts on the variable $q$. The von Neumann equation generalizes  the Schr\"odinger equation (\ref{SE}). 

As before, $W$-QM with just the von Neumann equation faces the quantum measurement problem. Below we write down  three solutions to the measurement problem with a fundamental density matrix. 

(1) \textbf{$W$-BM.} In addition to the density matrix $W$ that evolves unitarily according to the von Neumann equation, there are actual particles that have precise locations in physical space, represented by $\mathbb{R}^3$. The particle configuration $Q = (Q_1, Q_2, ... , Q_N) \in \mathbb{R}^{3N}$ follows the guidance equation (written for the $i$-th particle):\footnote{This version of the guidance equation is first proposed by \cite{bell1980broglie}, then discussed for a fundamental density matrix in \cite{durr2005role}.} 
\begin{equation}\label{WGE}
\frac{dQ_i}{dt} = \frac{\hbar}{m_i} \text{Im} \frac{\nabla_{q_{i}}  W (q, q', t)}{ W (q, q', t)} (q=q'=Q),
\end{equation}
Moreover, the initial particle distribution is given by the density-matrix version of the quantum equilibrium distribution: 
\begin{equation} \label{WQEH}
P(Q(t_0) \in dq) = W (q, q, t_0) dq.
\end{equation}
The system is also equivariant: if the probability distribution holds at $t_0$, it holds at all times.\footnote{Equivariance holds because of the following continuity equation: 
$$\frac{\partial  W(q,q,t) }{\partial t} = -\text{div} (  W(q, q, t) v),$$ where $v$ denotes the velocity field generated via (\ref{WGE}). See \cite{durr1992quantum, durr2005role}. We will discuss this equation in more detail in \S3.} 

(2) \textbf{$W$-GRW.} As before, there are several versions of $W$-GRW theories.  In the first one, $W$-GRW0, the fundamental ontology consists only in terms of the universal density matrix. The density matrix typically obeys the von Neumann equation, but the linear evolution is interrupted randomly (with rate $N\lambda$, where $N$ is the number of particles and $\lambda$ is a new constant of nature of order $10^{-15}$ s$^{-1}$) by collapses:\footnote{I am indebted to Roderich Tumulka for the following three equations and to Sheldon Goldstein and Matthias Leinert for helpful discussions.}
\begin{equation}\label{collapse}
W_{T^+} = \frac{\Lambda_{I_{k}} (X)^{1/2} W_{T^-} \Lambda_{I_{k}} (X)^{1/2}}{\text{tr} (W_{T^-} \Lambda_{I_{k}} (X)) }
\end{equation}
with $I_k$ uniformly distributed in the $N$-element set of particle labels and $X$ distributed by the following probability density:
\begin{equation}\label{center}
\rho(x) = \text{tr} (W_{T^-} \Lambda_{I_{k}} (x)),
\end{equation}
where the collapse rate operator is defined as before in (\ref{collapserate}). 

As before, we can add primitive ontology to the GRW theory. For $W$-GRWm, the version with a mass-density ontology, the mass-density is defined as a function of variables of physical spacetime: 
\begin{equation}\label{Wmxt}
m(x,t) = \text{tr} (M(x) W(t)),
\end{equation}
where $M(x)$ is the mass-density operator as defined after (\ref{mxt}). This allows us to determine the mass-density ontology at time $t$  via $W(t)$. In $W$-GRWf, the fundamental ontology includes $F$, as given in (\ref{F}), a collection of spacetime events, the spatial components of which are the centers of the spontaneous collapses of $W(t)$.

(3) \textbf{$W$-EQM.} There are two versions of Everettian theories with a fundamental density matrix. In the first one, $W$-EQM0, the fundamental ontology consists of only the universal density matrix. The density matrix always and exactly obeys the von Neumann equation. 

The Everettian theory can also be given a primitive ontology, such as a mass-density function. By using (\ref{Wmxt}), we define a mass-density function $m(x,t)$ from the universal density matrix. That will be part of the fundamental ontology. We thus obtain $W$-EQMm. 

The ontology problem and the probability problem for Everett take on a different form, and they seem more challenging in  $W$-EQM theories than in the $\Psi$-EQM theories. In general, the universal density matrix can be decomposed into wave functions that have overlapping supports in the configuration space. Even if each wave function, by decoherence, giving rise to a branching structure, it is not clear whether there will be such a branching structure for the collection of overlapping wave functions. By the linearity of the von Neumann equation, we know that each wave function will evolve linearly without interference from the other wave functions. So there is still independence, but it is not at all clear how to recover the talks about branches in $W$-EQM theories. Since the branching structure is crucial for the attempted solutions to the ontology problem and the probability problem in  $\Psi$-EQM theories, the strategies will need to be significantly modified to apply to the  $W$-EQM theories. It is interesting that the extension from $\Psi$-BM to $W$-BM requires no such changes, as the analysis of  probability and ontology in a Bohmian universe remains largely the same.

\subsubsection{$W_{PH}$-QM}

Once we have the $W$ theories, we can proceed to consider how to implement the low-entropy initial condition in them. First, it is important to notice that in such theories $W$ is the quantum statistical microstate. So our discussions of Boltzmannian quantum statistical mechanics in terms of $\psi$ will need to be adapted for $W$. Here are the key changes: 
\begin{itemize}
\item Being in a macrostate: typically, a density matrix is in a superposition of macrostates and is not entirely in any one of the macrospaces. However, we can make sense of situations where $W$ is very close to a macrostate $\mathscr{H}_\nu$: 
\begin{equation}\label{Wclose}
\text{tr} (W I_{\nu})  \approx 1,
\end{equation}
where $I_{\nu}$ is the projection operator onto $\mathscr{H}_{\nu}$. This means that almost all of $W$ is in $\mathscr{H}_{\nu}$. In this situation, we say that $W$ is in macrostate $\mathscr{H}_{\nu}$. 
\item Thermal equilibrium: typically, there is a dominant macro-space $\mathscr{H}_{eq}$ that has a dimension that  is almost equal to D: 
\begin{equation}
\frac{\text{dim} \mathscr{H}_{eq}}{\text{dim} \mathscr{H}} \approx 1.
\end{equation}
A system with density matrix $W$ is in equilibrium if  $W $ is very close to $\mathscr{H}_{eq}$ in the sense of (\ref{Wclose}):  $\text{tr} (W I_{eq})  \approx 1$.
\item Boltzmann entropy: the Boltzmann entropy of a quantum-mechanical system with density matrix $W$ that is very close to a macrostate $\nu$ is given by:
\begin{equation}\label{Boltzmann}
S_B (W) = k_B \text{log} (\text{dim} \mathscr{H}_\nu ),
\end{equation}
for which $W$ is in  macrostate $\mathscr{H}_\nu$  in the sense of (\ref{Wclose}). 
\end{itemize}

Next, let us consider how to adapt the Past Hypothesis for density matrices. Recall that, for $\Psi_{PH}$-QM,  we use (\ref{PH}) to constrain the initial wave functions: every initial wave function is entirely contained in the Past Hypothesis subspace $\mathscr{H}_{PH}$. Similarly, for density-matrix theories, we will propose that every initial density matrix is entirely contained in the Past Hypothesis subspace: 
\begin{equation}\label{WPH}
\text{tr} (W(t_0) I_{PH}) =1  \text{ , } \text{dim} \mathscr{H}_{PH} \ll \text{dim}\mathscr{H}_{eq} \approx \text{dim} \mathscr{H}
\end{equation}
where $I_{PH}$ is the projection operator onto the Past Hypothesis subspace, which is consistent with our notation so far. 

Given the $W$-version of the Past Hypothesis, every initial density matrix will be contained in the low-entropy subspace $\mathscr{H}_{PH}$. By time-reversal invariance, we know that there exist anti-entropic density matrices $\mathscr{H}_{PH}$. Nonetheless, it is reasonable to expect that these anti-entropic density matrices are atypical. However, unlike the situation for $\Psi_{PH}$ theories, it is far from clear whether there is any canonical probability (or typicality) measure for density matrices in a subspace.  If there is, it is   unlikely to be as natural and simple as the surface area measure on the unit sphere.

\subsubsection{$W_{IPH}$-QM}

Surprisingly, although there does not seem to be a canonical probability measure in $W_{PH}$-theories, there is a natural choice for the initial density matrix. Recall that, for every vector space $V$, there is a projection operator $I_V$ such that it is idempotent and takes every vector in $V$ to itself. Moreover, the projection operator is unique. In some sense, the projection operator encodes all the information about $V$. For example, when we say that a wave function is entirely contained or almost entirely contained in some subspace $\mathscr{H}_\nu$, we express the statement in terms of the projection $I_\nu$. So the choice of the projection is canonical given the choice of any vector space. 

The Past Hypothesis picks out a particular subspace $\mathscr{H}_{PH}$. It is canonically associated with its projection $I_{PH}$. In matrix form, it is the diagonal matrix that has a $k\times k$ identity block, with $k = \text{dim}{\mathscr{H}_{PH}}$, and zero everywhere else. There is a natural density matrix associated with $I_{PH}$, namely the normalized projection $\frac{I_{PH}}{\text{dim} \mathscr{H}_{PH}}$. Hence, we have picked out the natural density matrix associated with the Past Hypothesis subspace. We propose that the initial density matrix is the normalized projection onto $\mathscr{H}_{PH}$:
\begin{equation}\label{IPH}
\hat{W}_{IPH}(t_0) = \frac{I_{PH}}{\text{dim} \mathscr{H}_{PH}}.
\end{equation}
We  call this postulate the \emph{Initial Projection Hypothesis} (IPH). Crucially, it is different from (\ref{PH}) and (\ref{WPH});  while IPH picks out a unique  quantum state given the Past Hypothesis, the other two permit infinitely many possible  quantum states inside the Past Hypothesis subspace. 

Remarkably, we no longer need a fundamental postulate about probability or typicality. We know that we can decompose a density matrix (non-uniquely) into a probability-weighted sum of pure states, and in this case we can decompose $\hat{W}_{IPH}(t_0)$ as an integral of pure states on the unit sphere of $\mathscr{H}_{PH}$ with respect to the  uniform probability distribution:
\begin{equation}\label{MacroW}
\hat{W}_{IPH} (t_0) = \int_{\mathscr{S}(\mathscr{H}_{PH})} \mu(d\psi) \ket{\psi} \bra{\psi}.
\end{equation}
If the B-Conjecture in \S2.1.2 is true, then $\mu$-most wave functions in $\mathscr{H}_{PH}$ will increase in entropy and eventually approach thermal equilibrium. Thus, almost all of $\hat{W}_{IPH} (t)$ will be in higher-entropy macrostates and eventually approach thermal equilibrium. Hence, assuming the B-Conjecture, we know that $\hat{W}_{IPH} (t)$ will be (in the sense of (\ref{Wclose})) in a higher-entropy macrostate or in a superposition of higher-entropy macrostates and will eventually be in thermal equilibrium. It is worth emphasizing that the probability measure is not fundamental in the theory for two reasons. First,  the fundamental density matrix is, in the first instance, not to be interpreted as a probabilistic mixture of wave functions. Second, the decomposition into probabilistic mixture is not unique; we could have chosen a discrete probability distribution that assigns equal weights to the vectors in $\{\ket{n}  \}$, an orthonormal basis of  $\mathscr{H}_{PH}$:
\begin{equation}\label{MacroW2}
\hat{W}_{IPH} (t_0) =  \sum_{n=1}^{k} \frac{1}{k} \ket{n} \bra{n}.
\end{equation}

\begin{table}
\centering
    \begin{tabular}{ | l | c | c | }
    \hline
    Theory & Possible Initial Quantum States &  Natural Probability Measure    \\ \hline
    $\Psi$-QM & no restrictions  & $\mu$; uniform   \\ \hline
     $\Psi_{PH}$-QM & any $\Psi(t_0)$ s.t. $\Psi(t_0) \in \mathscr{H}_{PH}$ & $\mu$; uniform   \\ \hline
     $\Psi_?$-QM &  unclear & unclear   \\ \hline
     $W$-QM & no restrictions & unclear   \\ \hline
    $W_{PH}$-QM & any $W(t_0)$ s.t. $\text{tr}(W(t_0) I_{PH}) = 1$ & unclear   \\ \hline
    $W_{IPH}$-QM & $\hat{W}_{IPH}(t_0) = \frac{I_{PH}}{\text{dim} \mathscr{H}_{PH}}$ & unnecessary   \\ \hline
    \end{tabular}
\caption{A summary of the quantum theories discussed in \S2.}
\end{table}

In this section, we have presented two types of quantum theories: theories with a fundamental wave function and theories with a fundamental density matrix. We have found that there is a significant difference between the two, namely, given the Past Hypothesis:

\begin{itemize}
\item There is a natural  probability (typicality) measure over initial wave functions, but there is no natural choice for the initial wave function. 
\item There does not seem to be a natural  probability (typicality) measure over initial density matrices, but there is a natural choice for the initial density matrix, which makes the probability measure unnecessary. 
\end{itemize}
We summarize the finding in Table 1.

\section{Empirical Equivalence}

The two types of theories, one given by a fundamental wave function $\Psi$ and the other given by a fundamental density matrix $W$, are intimately related to each other. The von Neumann equation  for the evolution of $W$ is a generalization of the Schr\"odinger equation of $\Psi$; the $W$ guidance equation is a generalization of the $\Psi$ guidance equation.   The same is true for the collapse equation  and the definition of primitive ontology in the GRW theories. However, there are also many differences. The $W$ theories not only allow universal pure states but also allow  universal mixed states, while the latter are impossible in the $\Psi$ theories. 

It is natural to wonder whether we have empirical grounds for distinguishing between the two types of theories. In particular, we may ask: can a theory with a universal mixed state be empirically equivalent to one with a universal pure state? In this section, we will show that the answer is yes. In \S3.1, we present a general theorem for Bohmian theories from which we obtain a corollary that $\Psi_{PH}$-BM is empirically equivalent to $W_{IPH}$-BM. We also show that the Everettian theories are also empirically equivalent and the GRW theories are empirically equivalent. In \S3.2, we provide a more concrete understanding of their empirical equivalence by focusing on the Bohmian theories. In particular, we obtain some notions about subsystems in $W$-BM in parallel to those in $\Psi$-BM, including effective density matrix, conditional density matrix, collapse, and effective collapse.  We derive the fundamental conditional probability formula for $W$-BM.

\subsection{General Argument}

\subsubsection{Bohmian Theories}
Let us now focus on $\Psi$-BM and $W$-BM. Regarding the question of their empirical equivalence, \cite{durr2005role} write: 

\begin{quote}
One may wonder whether one can decide empirically between Bohmian mechanics and
$W$-Bohmian mechanics, or, in other words, whether one can determine empirically in a
universe governed by $W$-Bohmian mechanics if the fundamental density matrix is [pure]. The question is delicate. We think that the answer is no, for the following reason:
compare a W-Bohmian universe with a Bohmian universe with a random wave function
such that the associated statistical density matrix equals the fundamental density
matrix of the W-Bohmian universe. Since an empirical decision, if it can be made at
time [$t$], would have to be based solely on the configuration [$Q_{t}$] at that time, and since
the distribution of [$Q_{t}$]  is the same in both situations, it seems that there cannot be a
detectable difference: A given [$Q_{t}$]  could as well have arisen from an appropriate wave
function from the random wave function ensemble as from the corresponding fundamental
density matrix. (\cite{durr2005role}, \S7.3, with ``$t_0$''  replaced by ``$t$'' to avoid ambiguity)
\end{quote}
Here, the comparison is between $\Psi$-BM with a statistical density matrix $W$ and $W$-BM with a fundamental density $W$. Any empirical differences between the two theories, at time $t$, will manifest as differences in the outcomes of experiments, which are grounded in the configuration of particles at $t$. However, the probability distribution of $Q(t)$ is the same on both theories, which means the typicality facts are the same on both theories. Hence, the two cannot be empirically distinguished.  In what follows, I will expand this argument by stating a criterion of empirical equivalence and proving a theorem using that criterion. 

\begin{description}
\item[Criterion for Empirical Equivalence:] Theories $A$ and $B$ are empirically equivalent if they assign the same probability to every outcome in every measurement.
\end{description}
From this general criterion, we can derive a special criterion for Bohmian theories. Since in a Bohmian universe every measurement apparatus is made out of particles with precise positions, every measurement boils down to a position measurement. So we just need the two theories to agree on probability of the particle configurations. 
\begin{description}
\item[Criterion for Empirical Equivalence of Bohmian Theories:] Bohmian theories $A$ and $B$ are empirically equivalent if at any time $t$, $P_A (Q_t \in dq) = P_B (Q_t \in dq)$.
\end{description}

\begin{theorem}
Let $W_R$-BM be the theory of ($W$, $Q$) such that $W$ evolves by (\ref{VNM}), $Q$ evolves by (\ref{WGE}) and satisfies (\ref{WQEH}); moreover, a particular  $W(t_0)$ is chosen. Let $\Psi_R$-BM be the theory of ($\Psi$, $Q$) such that $\Psi$  evolves by (\ref{SE}), $Q$ evolves by (\ref{GE}) and  satisfies (\ref{QEH}); moreover, $\Psi(t_0)$ is chosen at random from a statistical ensemble represented by the density matrix $W(t_0)$. $W_R$-BM and $\Psi_R$-BM are empirically equivalent. 
\end{theorem}

Remark: $W_R$-BM is a more restrictive version of $W$-BM. $W_{IPH}$-BM corresponds to a particularly natural choice of $W(t_0)$. $\Psi_R$-BM is a version of $\Psi$-BM with an initial probability distribution over wave functions. $\Psi_{PH}$-BM corresponds to a particularly natural choice of the probability measure given the Past Hypothesis. 

Proof: At $t_0$, the particle configurations are distributed as follows: 


\begin{itemize}
\item  $W$-BM: $P_{W_R-BM} (Q_0 \in dq) = W(q, q, t_0) dq$.
\item  $\Psi_R$-BM: $P_{\Psi_R-BM} (Q_0 \in dq) = W(q, q, t_0) dq$.
\end{itemize}
$P_{W_R-BM}$ follows from (\ref{WQEH}). $P_{\Psi_R-BM}$ follows from (\ref{QEH}) and the definition of the density matrix $W$: 
\begin{equation}\label{ANYW}
\hat{W} = \int \mu(d\psi) \ket{\psi} \bra{\psi},
\end{equation}
where $\mu$ is the appropriately chosen measure for the statistical ensemble. In $\Psi_R$-BM, $\Psi$ satisfies the Schr\"odinger equation (\ref{SE}), from which we obtain the continuity equation: 
\begin{equation}\label{continuity}
\frac{\partial \rho}{\partial t} = -\text{div} ( J ). 
\end{equation}
We choose $\rho=|\psi(q, t)|^2$ and $v = J/\rho$. Then $|\psi(q, t)|^2$ is equivariant: 
\begin{equation}\label{Equivariance}
\frac{\partial |\psi(q, t)|^2 }{\partial t} = -\text{div} (  |\psi(q, t)|^2 v^\Psi),
\end{equation}
where $v^\Psi$ denotes the velocity field generated by (\ref{GE}). Let us integrate both sides of (\ref{Equivariance}) with respect to $\mu$, then we have:
\begin{equation}\label{AvEquivariance}
\frac{\partial \langle|\psi(q, t)|^2\rangle_\mu }{\partial t} = -\text{div} (  \langle |\psi(q, t)|^2 v^\Psi\rangle_\mu ),
\end{equation}
and equivalently,
\begin{equation}\label{WEquivariance}
\frac{\partial W(q,q,t) }{\partial t} = -\text{div} (  W(q, q, t) v^W),
\end{equation}
where $W(q,q,t) = \int \mu(d\psi) |\psi(q, t)|^2 = \langle \rho\rangle_\mu $ and $v^W = \langle J \rangle_\mu / \langle \rho\rangle_\mu$ denotes the velocity field generated by (\ref{WGE}). Hence $W(q,q,t)$ is also equivariant. Therefore, the probability formulae for $W_R$-BM and $\Psi_R$-BM will remain of the same form at any time:
\begin{itemize}
\item  $W_R$-BM: $P_{W_R-BM} (Q_t \in dq) = W(q, q, t) dq$.
\item  $\Psi_R$-BM: $P_{\Psi_R-BM} (Q_t \in dq) = W(q, q, t) dq$.
\end{itemize}
Therefore, $W_R$-BM and $\Psi_R$-BM are empirically equivalent.  $\square$

It follows from the previous theorem that: 
\begin{corollary}
$W_{IPH}$-BM and $\Psi_{PH}$-BM are empirically equivalent. 
\end{corollary}


\subsubsection{Everettian and GRW Theories}

For the Everettian theories, we can draw the same conclusion about the respective theories. Let $W_R$-EQM be the theory of $W$ such that $W$ evolves by (\ref{VNM}); moreover, a particular  $W(t_0)$ is chosen. Let $\Psi_R$-EQM be the theory of $\Psi$ such that $\Psi$  evolves by (\ref{SE}); moreover, $\Psi(t_0)$ is chosen at random from a statistical ensemble represented by the density matrix $W(t_0)$. $W_R$-EQM and $\Psi_R$-EQM are empirically equivalent.  

The reason is that they assign the same probability to every outcome in every experiment. Let $A$ be a self-adjoint operator corresponding to some observable such that its spectrum may not be completely discrete. Suppose  that its spectral measure is given by $\mathscr{A}$, a projection-valued measure. Then the probability that at time $t$, $x$ the outcome of the measurement will be within some measurable set $M$ is:  
\begin{itemize}
\item  $W_R$-EQM: $P_{W_R-EQM} (x \in M) = \text{tr} ( W_t\mathscr{A}(M))$.
\item  $\Psi_R$-EQM: $P_{\Psi_R-EQM} (x \in M) = \text{tr} ( W_t\mathscr{A}(M) )$.
\end{itemize}
Hence, they are empirically equivalent. 

For the GRW theories, the $W_R$ and $\Psi_R$ versions with different choices of the primitive ontology ($m$ or $f$) are also empirically equivalent to each other (with the arrows denoting empirical equivalence):

\begin{equation}\label{diag}
\xymatrix{
  \Psi_R\text{-GRWf} \ar[r] \ar[d]   & W_R\text{-GRWf} \ar[d] \ar[l] \\
  \Psi_R\text{-GRWm} \ar[r]  \ar[u] & W_R\text{-GRWm} \ar[u] \ar[l]
}
\end{equation}

 First, \cite{allori2013predictions} show that $W_R$-GRWf is physically equivalent to $\Psi_R$-GRWf, which implies that they are empirically equivalent. Their argument focuses on the joint distribution of all flashes. 
\begin{equation}
P_{W_R-GRWf} (F \in S) = \text{tr} (G_0(S) W_0) = P_{\Psi_R-GRWf} (F \in S),
\end{equation}
where $S$ is any set of flash histories, $G_0(\cdot)$ is the suitable positive-operator valued measure  (POVM) governing the distribution of the flashes in $\Psi$-GRWf, and $W_0$ is the initial density matrix. 

Second, \cite{goldstein2012quantum}  show that $\Psi$-GRWf is macro-history equivalent to $\Psi$-GRWm, which implies that they are empirically equivalent. They consider a macroscopic system corresponding to a pointer that can point to either position 1 or position 2 at time $t$. The wave function of the system has the form $\Psi_t = c_1 \Phi_1 + c_2 \Phi_2$, where $|c_1|^2+|c_2|^2=1$ and $\Phi_i$ is concentrated on configurations for which the pointer points to position 1. Suppose, from the perspective of $\Psi$-GRWf,  most flashes are in position 1. Then $|c_1|^2 \gg |c_2|^2 $, which means $m(1,t) \gg m(2,t)$. Hence, from the perspective of $\Psi$-GRWm, matter is concentrated in position 1. Now,  suppose, from the perspective of $\Psi$-GRWm, matter is concentrated in position 1. Then $m(1,t) \gg m(2,t)$, which means $|c_1|^2 \gg |c_2|^2 $. Hence, from the perspective of $\Psi$-GRWm, with overwhelming probability, most flashes are  in position 1. 

Third, we can follow the previous argument and show that $W_R$-GRWf  is macro-history equivalent to $W_R$-GRWm, which implies that they are empirically equivalent.  Again, let us consider a macroscopic system corresponding to a pointer  that can point to either position 1 or position 2 at time $t$. The density matrix of the system has the form $W_t = c_1 W_1 + c_2 W_2 + c_3 W_3$, where $c_1+c_2=1$, $W_1$ and $W_2$ are concentrated on configurations for which the pointer points to position 1 and to position 2 respectively, and $W_3$ have disjoint support from both $W_1$ and $W_2$.  Suppose, from the perspective of $W_R$-GRWf,  most flashes are in position 1. Then $c_1 \gg c_2  $, which means $c_1  \gg c_2  $. From the perspective of $W_R$-GRWm, matter is concentrated in position 1. The other direction is similar as above. 

Therefore, by the transitivity of empirical equivalence relation, we can conclude that all four versions of GRW theories in (\ref{diag}) are empirically equivalent.

\subsection{Subsystem Analysis of $W$-BM}

For Bohmian theories, the previous general argument and the corollary suffice to show that the relevant types of Bohmian theories can be empirically equivalent. However, in actual practice, it is rare to use the quantum state of the universe or to use the probability distribution over universal quantum states.  For the wave function versions, \cite{durr1992quantum} have provided a subsystem analysis that is used to clarify and justify the quantum equilibrium hypothesis (Born rule distribution). It would be illuminating to see that the usual Bohmian subsystem analysis carries over to the density-matrix theories, which will also shed further light on their empirical equivalence. However, it is beyond the scope of this paper to conduct a thorough statistical analysis of the quantum equilibrium hypothesis. Nevertheless, we will closely follow their strategies to define conditional density matrix, effective density matrix, and effective collapse, which we will then use to derive the fundamental conditional probability formula. These will show that it is possible to derive a $W$-version of the quantum equilibrium hypothesis using the method of \cite{durr1992quantum}. 

\subsubsection{An Example}

We begin by considering a simple example. Suppose the universal configuration is split into $Q=(X,Y)$, where $X$ is the configuration of some subsystem of interest and $Y$ is that of the environment.  Consider three quantum states below where $x$ and $y$ denote generic variables for the subsystem and the environment:

\begin{equation}\label{W1}
W_1(x,y, x', y') =   \Psi_1(x, y) \Psi_1^*(x', y').
\end{equation}

\begin{equation}\label{W2}
W_2(x,y, x', y') =   \Psi_2(x, y) \Psi_2^*(x', y').
\end{equation}

\begin{equation}\label{W3}
W_3(x,y, x', y') =   \frac{1}{2} W_1(x,y, x', y') + \frac{1}{2} W_2(x,y, x', y').
\end{equation}

Suppose further that $\Psi_1$ adequately describes a situation after a measurement: 

\begin{equation}\label{psi}
 \Psi_1(x, y) = \psi (x) \phi_1(y) + \Psi_1^\bot (x,y),
\end{equation}
such that $\phi_1$ and $\Psi_1^\bot$ have macroscopically disjoint $y$-supports (which also means that $\phi_1^*$ and $\Psi_1^{\bot*}$ have macroscopically disjoint $y'$-supports) and 
\begin{equation}\label{support1}
Y \in \text{supp } \phi_1,
\end{equation}
from which we can deduce that $Y \in \text{supp } \phi_1^*$. 

Similarly for $\Psi_2$, suppose that it also adequately describes the same measurement: 

\begin{equation}\label{phi}
 \Psi_2(x, y) = \psi (x) \phi_2(y) + \Psi_2^\bot (x,y),
\end{equation}
such that $\phi_2$ and $\Psi_2^\bot$ have macroscopically disjoint $y$-supports (which also means that $\phi_2^*$ and $\Psi_2^{\bot*}$ have macroscopically disjoint $y'$-supports) and 
\begin{equation}\label{support1}
Y \in \text{supp } \phi_2,
\end{equation}
from which we can deduce that $Y \in \text{supp } \phi_2^*$.

Let us now expand $W_1$: 
\begin{align}\label{W1+}
 W_1(x,y, x', y') &= \Psi_1(x, y) \Psi_1^*(x', y')  \nonumber \\
						& =  (\psi (x) \phi_1(y) + \Psi_1^\bot (x,y)) (\psi^* (x') \phi_1^*(y') + \Psi_1^{\bot*} (x',y')) \nonumber \\
						& =  \psi (x) \phi_1(y) \psi^* (x') \phi_1^*(y') + [\psi (x) \phi_1(y) \Psi_1^{\bot*} (x',y') + \nonumber \\
						&   \qquad { } \Psi_1^\bot (x,y)\psi^* (x') \phi_1^*(y') +  \Psi_1^\bot (x,y)\Psi_1^{\bot*} (x',y')] \nonumber \\
						&  = M_1(x,y,x',y') +  W_1^\bot(x,y,x',y'),
\end{align}
where 
\begin{equation}\label{M1}
M_1(x,y, x', y')  =   \psi (x) \phi_1(y) \psi^* (x') \phi_1^*(y'),
\end{equation}
and 
\begin{equation}\label{M1}
W_1^\bot(x,y, x', y')  =   \psi(x) \phi_1(y) \Psi_1^{\bot*} (x',y') + \Psi_1^\bot (x,y)\psi^* (x') \phi_1^*(y') +  \Psi_1^\bot (x,y)\Psi_1^{\bot*} (x',y').
\end{equation}
Since $\phi_1$ and $\Psi_1^\bot$ have macroscopically disjoint $y$-supports, and  $\phi_1^*$ and $\Psi_1^{\bot*}$ have macroscopically disjoint $y'$-supports, we know that $M_1(x,y, x', y') $ and $W_1^\bot(x,y, x', y')$ have macroscopically disjoint $(y,y')$-supports.

Similarly, 
\begin{align}\label{W2+}
 W_2(x,y, x', y') 		& =  \psi (x) \phi_2(y) \psi^* (x') \phi_2^*(y') + [\psi(x) \phi_2(y) \Psi_2^{\bot*} (x',y') + \nonumber \\
						&   \qquad { } \Psi_2^\bot (x,y)\psi^* (x') \phi_2^*(y') +  \Psi_2^\bot (x,y)\Psi_2^{\bot*} (x',y')] \nonumber \\
						&  = M_2(x,y,x',y') +  W_2^\bot(x,y,x',y'),
\end{align}
Similarly,  $M_2(x,y, x', y') $ and $W_2^\bot(x,y, x', y')$ have macroscopically disjoint $(y,y')$-supports.

Thus, we can  expand $W_3$ as follows:
\begin{align}\label{W3+}
 W_3(x,y, x', y') & = \frac{1}{2} W_1(x,y, x', y') + \frac{1}{2} W_2(x,y, x', y')  \nonumber \\
						& = \frac{1}{2}  [M_1(x,y,x',y')  +  M_2(x,y,x',y') ]   \\ 
				& +  \frac{1}{2} [W_1^\bot(x,y,x',y') + W_2^\bot(x,y,x',y')] \nonumber\\
				& = M_3(x,y, x', y') + W_3^\bot (x,y, x', y'),
\end{align}
where  
\begin{align}\label{M3}
M_3(x,y, x', y')  & =   \frac{1}{2}  [M_1(x,y,x',y')  +  M_2(x,y,x',y') ] \nonumber \\
& = \frac{1}{2} [\psi (x) \phi_1(y) \psi^* (x') \phi_1^*(y') + \psi (x) \phi_2(y) \psi^* (x') \phi_2^*(y') ]
\end{align}
\begin{equation}\label{M1}
W_3^\bot(x,y, x', y')  =   \frac{1}{2} [W_1^\bot(x,y,x',y') + W_2^\bot(x,y,x',y')] .
\end{equation}
 In general, $M_3(x,y, x', y')$ and $W_3^\bot (x,y, x', y')$ can have overlapping $(y,y)$-supports. However,  situations after measurements are quite special. Suppose (\ref{psi}) and (\ref{phi}) describe two wave functions after some experiment on the $x$-subsystem such that both wave functions are compatible with our observations. Then it is reasonable to expect that  $\text{supp } \phi_1$ and $\text{supp }  \phi_2$ do not differ by a macroscopically significant amount. In that case, we can infer that  $\phi_2$ and $\Psi_1^\bot$ have macroscopically disjoint $y$-supports and that $\phi_1$ and $\Psi_2^\bot$ have macroscopically disjoint $y$-supports. Therefore, in this case, $M_3$ and $W^\bot$ have macroscopically disjoint $(y,y)$-supports. 


Now, let us consider $M_3$. It has a product form: 
\begin{equation}\label{M3}
M_1(x,y, x', y')  =  \psi (x) \psi^* (x') [\phi_1(y) \phi_1^*(y') +\phi_2(y) \phi_2^*(y')  ]  = \rho(x,x') \gamma(y,y')
\end{equation}
 The effective density matrix of the subsystem is $\rho(x,x')$. Moreover, $\gamma(y,y')$ and $W_3^\bot$ have macroscopically disjoint $(y,y)$-supports by the previous argument. Hence, even if the universe is in a mixed state $W_3$, the subsystem  can nonetheless be in a pure state  $\rho(x,x')$. In this case, the effective density matrix $\rho$ corresponds to the wave function $\psi$. 

Now, let us define the conditional density matrix of the $x$-system as follows:
\begin{equation}\label{Wcond}
 w(x, x')^Y  =  W(x, Y, x', Y),
\end{equation}
where we identify quantum states related by a non-zero constant factor. In the case of $W_3$, the conditional density matrix is: 
\begin{equation}\label{W3cond}
 w_3(x, x')^Y  =  W_3(x, Y, x', Y) = M_3(x, Y, x', Y) + 0  = \rho(x,x') \gamma(Y,Y) = C \rho(x,x').
\end{equation}
where $C$ is a real number that can be taken care of by normalization. Here we have used $W_3^\bot (x,Y, x', Y)=0$, since we know that $Y$ is not in the union of the $y$-supports of $\Psi_1^{\bot}$ and $\Psi_2^{\bot}$. 

\subsubsection{General Statements}

Motivated by the previous example, we propose the following general statements about $W$-BM subsystems. 

(1) \textbf{Splitting.} For any given subsystem of particles we have a splitting:
\begin{equation}\label{splitting}
q=(x,y),
\end{equation}
with $x$ the generic variable for the configuration of the subsystem and $y$ the generic variable for the configuration of the environment, i.e. the complement of the subsystem. (\ref{splitting}) provides a splitting of the actual configuration into two parts:
\begin{equation}\label{actualsplitting}
Q=(X,Y).
\end{equation}
So we can write the universal density matrix in terms of $W = W(x,y, x',y') $. 

(2) \textbf{Effective density matrix.} 
The subsystem corresponding to the $x$-variables has an \emph{effective density matrix } (at a given time) if the universal density matrix $W(x,y,x',y')$ and the actual configuration $Q=(X,Y)$ (at that time) satisfy: 
\begin{equation}\label{Weffective}
W(x,y,x',y') = \rho(x,x')\gamma(y,y') + W^\bot (x,y, x', y'), 
\end{equation}
such that  $\rho$ is a pure state, 
$\gamma(y,y')$ and $ W^\bot (x,y, x', y')$ have macroscopically disjoint $(y,y)$-supports,  and
\begin{equation}\label{supp}
(Y,Y) \in \text{supp } \gamma(y,y'),
\end{equation} 
In this case, the effective density matrix of the subsystem is $\rho(x,x')$. We expect that our definition of effective density matrix coincides with the usual practice of the quantum formalism for assigning pure states to systems, whenever the latter does assign pure states.\footnote{It does not seem to have a natural extension to  situations where both $\rho$ and $\gamma$ are mixed states.  I am assuming, as do \cite{durr1992quantum}, that standard quantum measurement recipe will render a pure state for the subsystem after measurement.}

(3) \textbf{Conditional density matrix.} The effective density matrix for a subsystem does not always exist. However, we can always define the \emph{conditional density matrix} in the following way:
\begin{equation}\label{conditional}
w(x,x')=W(x,Y,x',Y).
\end{equation}
Here we identify quantum states differing by a constant factor. Given the definition of the velocity (\ref{WGE}), the velocity field of the $x$-system will be given by its conditional density matrix. However, the conditional wave function does not always evolve in a unitary way, because of the interactions between the subsystem and the environment. However, when the system and the environment are suitably decoupled, such as after measurement when (\ref{Weffective}) and (\ref{supp}) are satisfied, then the conditional wave function becomes the effective wave function and will obey its own von Neumann equation. 

(4) \textbf{Collapse and effective collapse.} When (\ref{Weffective}) and (\ref{supp}) are satisfied, we can  neglect, for all practical purposes, $W^\bot(x,y,x',y')$. The configuration will be carried by the relevant part of the universal density matrix---$\rho(x,x')\gamma(y,y')$---into the future, without much interference from the other parts contained in  $W^\bot(x,y,x',y')$. In this case, we can say that during measurement,  the universal density matrix has undergone an effective collapse from $W_{t^-}$ to $W_{t^+}=\rho(x,x', t^+)\gamma(y,y',t^+)$. However, from the point of view of the subsystem density matrix, represented by the conditional density matrix $w(x,x',t)$, there is a real discontinuous change to  $w(x,x',t^+)$. Hence, the subsystem density matrix has undergone a genuine collapse, which is consistent with the prescriptions of textbook quantum mechanics. As in the $\Psi$-BM, the subsystem analysis of $W$-BM provides an explanation for the usefulness of talking about collapses even in a universe fundamentally governed by unitary dynamics. 

(5) \textbf{The Fundamental Conditional Probability Formula.} By equivariance (\ref{WEquivariance}) the distribution of $Q_t$ is always given by $W(q,q,t)$. By (\ref{conditional}), at time $t$, for the conditional probability distribution of the configuration of a subsystem $X_t$ given the actual configuration of the environment $Y_t$, we have the fundamental conditional probability formula for $W$ theories: 
\begin{equation}\label{conditionalprob}
P(X_t \in dx | Y_t) = w (x,x,t) dx,
\end{equation}
where $w (x,x',t) = w(x,x',t)^{Y_t} $ is the conditional density matrix of the subsystem at time $t$. Similar to the situation in $\Psi$-BM, the configurations $X_t$ and $Y_t$ are conditionally independent given the density matrix $w (x,x',t)$. As in the $\Psi$-BM situation, we expect  that we can  extract the entire empirical statistical content, including an analogous ``principle of absolute uncertainty,''  from the fundamental conditional probability formula. However, we will not attempt to carry out the task of a complete statistical analysis here. It is reasonable to expect that what we have provided above should be sufficient for that task.

\section{Discussions}

In \S2, we introduced  quantum theories with a fundamental density matrix. In \S3, we found that a certain class  of $W$ theories are empirically equivalent to a certain class of $\Psi$ theories.  In this section, I discuss some theoretical payoffs of $W$ theories and present some open questions for future research. 

\subsection{Theoretical Payoffs}

(1) \textbf{A natural initial condition.} Each version of $\Psi$-QM can be viewed as a special class of $W$-QM where the fundamental density matrix is pure. Every possibility of $\Psi$-QM is a possibility of $W$-QM, but not vice versa. The latter allows many more possibilities with mixed states. If $\Psi$-QM is empirically adequate, why should we be interested in a theory with redundant possibilities? 

One might appeal to the (controversial) paradox about black hole information loss. Suppose the universe consists in two systems that are entangled.  If we throw one system into an evaporating black hole, then the universal quantum state, even if it started as a pure state, will become mixed after the evaporation. So we seem to be forced to consider the possibility of universal mixed states. However, it is unclear to me whether $W$-QM can help with the paradox, since the pure to mixed transition is not allowed in any version of $W$-QM we considered in \S2.2.\footnote{Perhaps it can be helped by a $W$-GRW theory with a Lindblad equation as the fundamental equation of density matrix evolution. That is formulated as Mm in \cite{goldstein2012quantum}.}

One reason we should be interested in  $W$-QM, I think, is that they can be nicely combined with the Past Hypothesis. As we saw in \S2.2.3, the Past Hypothesis subspace suggests a natural choice of the initial density matrix $W_{IPH}(t_0)$. We thus obtain $W_{IPH}$-QM. Given a choice of the Past Hypothesis subspace, there is only one initial quantum state allowed by $W_{IPH}$-QM---the normalized projection. This is highly unusual in theoretical physics. Theories we usually consider allow for (infinitely) many initial states given an initial macrostate: Newtonian mechanics, Maxwellian electrodynamics, $\Psi$-QM, $\Psi_{PH}$-QM, $W$-QM,  and $W_{PH}$-QM. All versions of $W_{IPH}$-QM, with the exception of $W_{IPH}$-BM, allow for only one initial state at $t_0$, given the Past Hypothesis macrostate. Even for $W_{IPH}$-BM, there is only one possible initial \emph{quantum} state. Of course, the Past Hypothesis macrostate can be arbitrary. Given a choice of macro-variables, the boundary of ``the'' Past Hypothesis subspace can be fuzzy. However, since the projection onto $\mathscr{H}_{PH}$ now plays both the macroscopic role of giving us the low-entropy boundary condition and the microscopic role of figuring in the microscopic dynamics, the choice of the Past Hypothesis subspace might becomes less arbitrary.\footnote{Given a unique choice of the Past Hypothesis subspace, there is only one choice of the initial quantum state. For $W_{IPH}$-Everettian theories, since the dynamics is deterministic, that means there is only one possible history of the universe. This could be case of \emph{strong determinism} in the sense of \cite{penrose1999emperor}. }

(2) \textbf{Reduction of statistical mechanical probabilities. }  Fundamental statistical probabilities arise as a measure of typicality over initial phase points in classical mechanics or over initial wave functions in quantum mechanics. It is necessary because not every initial phase point and not every initial wave function will evolve to states of higher entropy. The anti-entropic initial states, however, have extremely small measure and are thus overwhelmingly unlikely. In the quantum case, the statistical mechanical probabilities are an additional source of randomness beyond the usual randomness of measurement outcomes (Born rule distribution). Hence, we have two sources of fundamental randomness in a theory such as $\Psi_{PH}$-QM. 

However, as we point out in \S2.2.3, fundamental statistical probabilities are no longer necessary in $W_{IPH}$-QM. Given the Past Hypothesis subspace, the theory allows only one initial quantum state---the normalized projection onto the subspace. If the B-Conjecture holds for typical wave functions, then it also holds for the normalized projection---it will, with certainty, evolve to states of higher entropy and eventually to thermal equilibrium. Hence, we no longer need statistical mechanical probabilities in addition to quantum mechanical ones. The only source of probabilities, in $W_{IPH}$-QM, are quantum mechanical probabilities, however they are to be understood in the end. 

(3) \textbf{Theoretical unity. } Theories with a fundamental density matrix also leads to an increased level of theoretical unity. In $W$-BM (and other more restrictive versions with PH or IPH), there is dynamical unity for the universal level and the subsystem descriptions. The universe is described by a density matrix;  the configuration of the universal system follows the $W$-BM guidance equations. The same is true for any subsystem: it is described by a conditional density matrix which guides the subsystem configuration via the $W$-BM guidance equation (\ref{WGE}). This is true even when we include spin. 

The situation is to be contrasted with the situation in $\Psi$-BM with spin, where the universe is in a pure state but a typical subsystem will not have a conditional wave function. Instead, it may only have a conditional density matrix, which guides the subsystem configuration not via the $\Psi$-BM guidance equation (\ref{GE}) but with the $W$-BM guidance equation (\ref{WGE}). 

For $W$-GRW and $W$-EQM, both the universe and the subsystem are described by density matrices. In $\Psi$-GRW and $\Psi$-EQM, the universe is described by a wave function but the subsystems are described by reduced density matrices which are typically mixed states. 

(4) \textbf{The nature of the quantum state. } Since the quantum state is defined on a high-dimensional space, it remains a puzzle in quantum foundations how to understand what it represents in the physical world. An attractive proposal, due to \cite{goldstein1996bohmian, goldstein2001quantum}, and \cite{goldstein2013reality}, is to regard it as nomological, i.e. on par with the fundamental laws of nature. The analogy is with the Hamiltonian function in classical mechanics: $H(p, q)$ can be understood as a simple description of the kinetic energy and pair-wise interactions among the point particles, so that it does not have to be represented as part of the material ontology but rather as on par with other dynamical laws of nature. To be nomological, $H$ satisfies four features: (a) it generates motion, (b) it is simple, (c) it is fixed by the theory, and (d) it does not represent things in the material ontology. 

The wave function in $\Psi$-BM satisfies (a), as it generates the velocity field on configuration space. However, generic wave functions are neither simple nor fixed by the theory. \cite{goldstein1996bohmian} observe that the wave function might be nomological in some theories of quantum gravity that satisfy the Wheeler-DeWitt equation. In those theories, the wave function can be regarded as stationary, which means it does not have time-dependence and probably have many symmetries. In that case, it could be quite simple. Given the simplicity, we can stipulate such a wave function explicitly in the Bohmian theory, which then can play the nomological role as the classical Hamiltonian. 

 $W_{IPH}$-BM also supports the nomological interpretation of the quantum state. At $t_0$, we have a quantum state $W_{IPH}(t_0)$ that is as simple and as unique as the Past Hypothesis subspace. Moreover, it has been suggested that the Past Hypothesis can be a law of nature. So we have another route to the simplicity of the initial quantum state---through the Initial Projection Hypothesis. Furthermore, any later quantum states can be written as products of the time-evolution operator and $W_{IPH}(t_0)$, both of which are simple.  In this sense, $W_{IPH}(t_0)$  generates motion in a simple way. This route to the nomological interpretation is novel because it does not depend on specific proposals about quantum gravity. 

\subsection{Open Questions}

$W$-QM give rise to several open questions. Here I mention two which I hope to address in future work. 

(1) \textbf{Non-normalizable quantum states. } $W_{IPH}$-QM requires the initial quantum state to be the normalized projection onto the Past Hypothesis subspace $\mathscr{H}_{PH}$. For technical reasons, we have assumed that $\mathscr{H}_{PH}$ is finite-dimensional, such that the projection can be normalized by $1/\text{dim}\mathscr{H}_{PH}$. If $\mathscr{H}_{PH}$ is infinite-dimensional, then the normalization will not work. Given an infinite dimensional $\mathscr{H}_{PH}$, we may not have a canonical density matrix as natural as $\frac{I_{PH}}{\text{dim}\mathscr{H}_{PH}}$. However, it is not hopeless, as we have learnt from quantum cosmology that there are many situations that we need to deal with non-normalizable quantum states. 

In cases where $\mathscr{H}_{PH}$ turns out to be infinite dimensional, it is still an option to take   the initial quantum state to be the projection $I_{PH}$. It is no longer a density matrix because it is not normalized. However, the dynamics is still  well-defined. The von Neumann equation (\ref{VNM}) governs the time evolution of $W(t)$, where $W(t_0) = I_{PH}$. Moreover, $W(t)$ can still give rise to a velocity field on configuration space via the $W$-Bohmian guidance equation (\ref{WGE}). The definition of the matter-density and flash ontology can be tricky given a non-normalizable quantum state. Let us call these theories $W_{IPH\infty}$-QM. The usual statistical analysis will not extend trivially to such theories. But perhaps something different can be applied. It is an open question what measure of typicality and what analysis of probability we can use when the universal quantum state is non-normalizable.

(2) \textbf{Ontology and probability of $W$-Everettian theories. } The Bohmian and GRW theories inherit no new problems about ontology or probability when we move from a wave-function theory to a density-matrix theory. However,  as we noted in \S2, new questions emerge for the $W$-Everettian theories. Since decoherence may not by itself be sufficient to show that there is an emergent branching structure in the fundamental density matrix, we cannot just use the standard arguments in $\Psi$-Everettian theories to justify the emergent ontology and probability. It is still an open question what and whether new techniques can be applied to solve these problems. If they can be solved, it would be technically interesting to compare the techniques in the two cases. If they cannot be solved, then it seems that the Everettian framework does not allow for universal mixed state on pain of being empirical inadequate. That would also be interesting news.

\section{Conclusion}

In this paper, we have looked at two types of quantum theories: one with a fundamental wave function and the other with a fundamental density matrix. We found that there is a crucial difference: on the first type of theories, there is a natural measure of probability (or typicality) over quantum states, but there is no natural choice for an initial quantum state, while the opposite is true on the second type---there is a natural choice for the initial quantum state but does not appear to be a natural measure of probability. We showed that they can nonetheless be empirically equivalent descriptions of the world. To that end, we gave some general arguments for agreement of measurement outcome statistics for Bohmian, GRW, and Everettian theories, and we introduced a novel subsystem analysis for the Bohmian theory. Finally, we suggested that there are some theoretical payoffs and open questions on the density-matrix approach. 

Is the universe in a pure state or a mixed state? We probably cannot know the answer  based on empirical grounds. However, if we would like to find a natural choice of the universal quantum state, we can easily do so if we allow the universe to be in a fundamental mixed state, with the initial quantum state given by the normalized projection onto the Past Hypothesis subspace.


\section*{Acknowledgement}

I am grateful for helpful discussions with David Albert, Sheldon Goldstein, Barry Loewer, Roderich Tumulka, and Nino Zangh\`i.


\bibliography{test}


\end{document}